# Tunneling conductance in strained graphene-based superconductor: Effect of asymmetric Weyl–Dirac fermions

**Bumned Soodchomshom**

Thailand Center of Excellence in Physics, Commission Higher on Education, Ministry of Education, Bangkok 10400, Thailand

**Abstract**

Based on the BTK theory, we investigate the tunneling conductance in uniaxially strained graphene-based normal metal (NG)/ barrier (I)/superconductor (SG) junctions. In the present model, we assume that by depositing the conventional superconductor on the top of the uniaxially strained graphene, normal graphene may turn to superconducting graphene with the Cooper pairs formed by the asymmetric Weyl-Dirac electrons, the massless fermions with direction-dependent velocity. The highly asymmetrical velocity, $v_y/v_x \gg 1$, may be created by strain in the zigzag direction near the transition point between gapless and gapped graphene. In the case of highly asymmetrical velocity, we find that the Andreev reflection strongly depends on the direction and the current perpendicular to the direction of strain can flow through the junction as if there was no barrier. Also, the current parallel to the direction of strain anomalously oscillates as a function of the gate voltage with very high frequency. Our predicted result is quite different from the feature of the quasiparticle tunneling in the unstrained graphene-based NG/I/SG conventional junction. This is because of the presence of the direction-dependent-velocity quasiparticles in the highly strained graphene system.

# 1. Introduction

Since graphene, a one-atomic-thick monolayer of graphite, has been first fabricated [1], it has become a new material with great potential for novel devices. Because of honeycomb-like lattice, electrons in grapheme mimic the massless relativistic Weyl–Dirac particles, with the Fermi velocity $v_F \sim 10^6$ m/s playing a role of the speed of light [2-4]. The energy spectrum E(k) of electrons in graphene exhibits the linear dispersion $E = \hbar v_F \sqrt{k_x^2 + k_y^2}$, obeying the spectrum of the massless relativistic particles. Electrons propagate in graphene with the constant velocity $v_F$ for all angles of incidence ie., $v_{x,y} = \partial E / \hbar \partial_{k_{x,x}} = v_F$. By having the carriers as massless relativistic fermions, graphene leads condensed matter into the world of quantum electrodynamics. In contrast to the Schrödinger-like electrons, massless relativistic electrons in graphene tunnel through a barrier without back reflection at the normal incidence, known as Klein paradox [5].

Specular Andreev reflection in graphene is one of the interesting effects appearing in graphene as a bridge between relativity and superconductivity, since graphene can be a superconductor by mean of proximity effect [6, 7]. Depositing conventional superconductor on the top of a graphene sheet leads the normal graphene (NG) to become superconducting graphene or graphene superconductor (SG) [6, 7]. Graphene superconductor fabricated by depositing Ti/Al (10/70nm) and Pt/Ta/Pt (3/70/3nm) on the top of graphene sheet give rise to the critical temperature of 1.3K [6] and 2.5 K [7], respectively. The relativistic Cooper pairs in such system are formed by Weyl-Dirac electrons with momentum $\vec{k}$ and spin up attracting to Weyl-Dirac electrons with momentum $-\vec{k}$ and spin down. The tunneling between normal graphene and superconducting graphene, a NG/SG junction, was first studied by Beenakker [8]. The combination between relativity and superconductivity leads to the specular Andreev reflection, occurring when the Fermi energy $E_F$ of NG is smaller than the biased energy eV. The conductance drops to zero at $eV=E_F$, the transition point between the retro and the specular Andreev reflections. Effect of the presence of the specular Andreev reflection in NG/SG junction also gives rise to a new aspect of the tunneling conductance which is quite different from that in the conventional N/S junction [9, 10]. In the case of the junction having a gate barrier, NG/I/SG junction [11-13], the conductance of the junction oscillates as a function of the gate voltage, also in contrast to the decaying behavior in the conventional N/I/S junction [9, 10].

Recently, electronic properties of the deformed graphene system have drawn much attention [14-22]. Remarkably, the locally strained graphene can induce a valley-dependent pseudo-vector potential perpendicular to the direction of stain, due to shifted valley-dependent Dirac point in the strained region [14-17]. This leads to the valley polarization, an important characteristic for valleytronics [14-17]. Also, a gigantic pseudo magnetic field greater than 300 Tesla resulting from the strongly deformed graphene was observed in graphene nanobubbles [18]. In the case of graphene being uniaxially strained, gapless graphene may turn to gapped graphene at the critical strain ($S_C$) [21-20]. Several groups predicted that energy gap in graphene may be opened up by applying tension in the zigzag direction [21, 22]. As in contrast to the electrons in the undeformed graphene system, for strain smaller than the critical value $S_C$, electrons in the strained-graphene exhibit asymmetric massless fermions governed by the asymmetric energy dispersion [22]

$$E = \hbar \sqrt{v_x^2 k_x^2 + v_y^2 k_y^2} , \qquad (1)$$

where $v_{x,y} = \partial E / \hbar \partial_{k_{x,x}} \neq v_F$ and $v_x \neq v_y$. The new effect of the direction-dependent velocity give rises to the asymmetrical transport property [19, 21]. The carriers of the strained graphene system with strain smaller than $S_C$ are governed by the two-dimensional asymmetric Weyl–Dirac Hamiltonian, as is given by [22, 23]

$$H = \hbar \begin{bmatrix} 0 & v_x k_x - iv_y k_y \\ v_x k_x + iv_y k_y & 0 \end{bmatrix}, \qquad (2)$$

where $v_x$ and $v_y$ depend on the geometry of the deformed graphene [22].

In this paper, we propose a model to show the effect of asymmetrical velocity $v_x \neq v_y$ of the massless fermions in the deformed graphene on the specular Andreev reflection in a NG/SG junction and the tunneling conductance in NG/I/SG junction. In the case of applying strain in the zigzag direction, the highly asymmetrical velocity $v_y \gg v_x (v_x \sim \text{small})$ is found at the strain approaching $S_C$. **By means of proximity-induced superconductor** [6, 7] when conventional superconductor is deposited on the top of strained graphene, superconductivity occurs due to the Cooper pairs formed by the asymmetric Weyl–Dirac fermions. Our work focuses on the effect of direction-dependent velocity on tunneling conductance of the system with highly asymmetric velocity, $v_y \gg v_x (v_x \sim \text{small})$. Using the Blonder–Thinkham–Klapwijk (BTK) theory [10], we show the new feature of the specular Andreev reflection and the conductance in the strongly deformed graphene NG/I/SG junction which are influenced by the effect of the asymmetric Weyl–Dirac fermions, instead of the symmetric Weyl–Dirac fermions in the undeformed graphene NG/I/SG conventional junctions [8, 11-13]. In our model we use the strain dependence of the geometry and hoping energies of grapheme, based on ref.21.

## 2.    Theory and formalism

### 2.1 Highly asymmetric Weyl–Dirac fermions in deformed graphene

Based on the tight-binding model, we straightforwardly calculate the Hamiltonian of free electrons in deformed graphene (see the deformed geometry in Fig.1a) by using the formalism [21-23], as given by

$$H = \begin{bmatrix} 0 & \phi(k = <k_x, k_y>) \\ \phi^*(k = <k_x, k_y>) & 0 \end{bmatrix}, (3)$$

where $\phi(k = <k_x, k_y>) = -(t_1 e^{i\vec{k}.\vec{\sigma}_1} + t_2 e^{i\vec{k}.\vec{\sigma}_2} + t_3 e^{i\vec{k}.\vec{\sigma}_3})$ and we let $t_1 = t_2 = t = t_3/\eta$ as the hoping energies with the asymmetric constant $\eta$. In the case of the deformed graphene, we have $\vec{\sigma}_1 = <L_x, -L_y>$, $\vec{\sigma}_2 = <-L_x, -L_y>$ and $\vec{\sigma}_3 = <0, c'>$. When applying strain **S** in the **armchair direction** (along the y-direction) by using the model of ref.21, we therefore have

$$L_x = (1 - pS)c\sqrt{3}/2, \ L_y = (c/2)(1 + S) \text{ and } c' = c(1 + S),$$

and in the **zigzag direction** (along the x-direction), we have

$$L_x = (1 + S)c\sqrt{3}/2, \ L_y = (c/2)(1 - pS) \text{ and } c' = c(1 - pS),$$

$$(4)$$

where the carbon-carbon distance c=0.142 nm and the Poisson's ratio p=0.165 are applied [21]. By expanding $\phi(k)$ around $k_x = k_D = \dfrac{1}{L_x}\cos^{-1}\left[-\eta/2\right]$ and $k_y = 0$ when $\eta < 2$ [22], we then have asymmetric Hamiltonian in eqn.(3) similar to eqn.(2) with the Eigen energy related to eqn.(1) in the form $E = \hbar\sqrt{v_x^2(k_x - k_D)^2 + v_y^2 k_y^2}$ [22]. The asymmetrical velocities when $\eta < 2$ can be obtained as

$$v_x = 2tL_x\sqrt{1 - \frac{\eta^2}{4}}/\hbar \text{ and } v_y = \eta t(L_y + c')/\hbar .$$

$$(5)$$

Note that gapless graphene may turn to gapped graphene when $\eta > 2$. In this work, we focus only on the case of gapless graphene. The carriers are massless fermions, and the condition of $\eta < 2$ is necessary. Using the hoping energies as decaying models $t = t_o e^{-3.37(\frac{|\vec{\sigma}_1|}{c} - 1)}$ and $\eta t = t_o e^{-3.37(\frac{|\vec{\sigma}_3|}{c} - 1)}$ [21] with $t_0$ being the hoping energy in the undeformed graphene, we find that the critical deformation point is found at strain of $S_C \sim 0.228855$ ($\eta = 2$) for strain applying in the zigzag direction as shown in Fig.(1b). In this numerical result, applying strain in the armchair direction gives rise to the gapless graphene, due to $\eta < 2$ for all strain. The effect of asymmetrical velocity when applying strain in the armchair direction yields very small $v_y/v_x \sim 0.6$ or $v_x/v_y \sim 1.67$. Unlike that in the case of applying strain in the zigzag direction, we have $v_y/v_x \to \infty$ when $S \to S_C$ which gives rise to the highly asymmetric velocity effect. Because of applying strain in the zigzag direction can cause the highly asymmetric velocity for fermions, in the next section, we focus this effect on the specular Andreev reflection and the tunneling conductance in the deformed graphene-based NG/I/SG junction.

## 2.2 Scattering process in deformed graphene-based NG/I/SG junctions

In this section, we investigate the tunneling conductance in NG/I/SG junctions in the case of graphene sheet being deformed. Graphene sheet is strained in the zigzag direction (see Fig.2). We focus on the two currents flow in x-direction (model in Fig.2a) and y-direction (model in Fig.2b). The junctions are biased by the potential V and the gate voltage $V_G$. As we have mentioned above, the Cooper pairs in the deformed graphene-based superconductor are assumed as formed by the asymmetric Weyl–Dirac electron with the spin $\uparrow$ and momentum $\vec{k}$ attracting to the asymmetric Weyl–Dirac electron with spin $\downarrow$ and momentum $-\vec{k}$. The BCS mean field Hamiltonian used to describe the electron field in SG for case of deformed graphene is

$$H_{BCS} \sim \int dxdy\hat{\psi}_\sigma^*(-i\hbar[v_x\sigma_x\partial_x + v_y\sigma_y\partial_y] + U(x,y))\hat{\psi}_\sigma + \int dxdy(\Delta^*(x,y)\hat{\psi}_\uparrow\hat{\psi}_\downarrow + \Delta(x,y)\hat{\psi}_\uparrow^*\hat{\psi}_\downarrow^*),$$

$$(6)$$

where $\hat{\psi}_\sigma$ and $\hat{\psi}_\sigma^*$ are the annihilation and creation field operators for the asymmetric Weyl-Dirac electron with spin $\sigma$, respectively. U(x,y) is the potential energy of a single electron, $\sigma_{x,y}$ are Pauli spin matrices, and $\Delta(x,y)$ is the superconducting order parameter. The wave equation, **asymmetric Weyl-Dirac Bogoliubov-de Gennes equation (BdG)**, related to the BCS-mean-field Hamiltonian in eqn.(6) is therefore given by

$$\begin{pmatrix} -i\hbar[v_x\sigma_x\partial_x + v_y\sigma_y\partial_y] + U(x,y) & \Delta(x,y) \\ \Delta^*(x,y) & i\hbar[v_x\sigma_x\partial_x + v_y\sigma_y\partial_y] - U(x,y) \end{pmatrix}\psi(x,y) = E\psi(x,y).$$

(7)

In eqn. (7), we have canceled the Dirac point shifting by assuming that graphene is homogeneously strained. Electrons have the same Dirac point for all regions so that we can obtained $(k_x - k_D, k_y) \rightarrow (k_x, k_y)$. Note that due to the effect of the Dirac point shifting, the case of the locally strained graphene is only considered as a pseudo vector potential in the strained region [14-16].

Let we first consider the scattering process due to the current parallel to the direction of strain ($I_x$). This model is illustrated in Fig.2a. In this case, the parallel (or conservation) momentum is the wave vector in the y-direction $k_y = k_{//}$. The superconducting order parameter with phase $\phi$ and the potential energy are defined as

$$\Delta(x,y) = \Delta e^{i\phi}\Theta(x - d),$$

and $\qquad U(x,y) = -E_F\Theta(-x - d) - (E_F + V_G)\Theta(x)\Theta(-x + d) - (E_F + U)\Theta(x - d),$

(8)

respectively. $E_F$, $V_G$ and $U$ are the Fermi energy in NG, the gate potential in the barrier (I) and the electrostatic potential in superconducting electrode SG, respectively. The wave solution to the BdG equation for each region is obtained as of the form

$$\psi(x < 0, y) = (\psi_{Ne+} + b\psi_{Ne-} + a\psi_{Nh+})e^{ik_{//}y},$$

$$\psi(0 < x < d, y) = (l\psi_{Ie+} + m\psi_{Ie-} + p\psi_{Ih+} + q\psi_{Ih-})e^{ik_{//}y},$$

and $\qquad \psi(d < x, y) = (c\psi_{Se+} + d\psi_{Sh-})e^{ik_{//}y},$

where $\qquad \psi_{Ne\pm} = \begin{pmatrix} 1, & \dfrac{E_F + E}{\pm\hbar v_x k_{Nx,e} - i\hbar v_y k_{//}}, & 0, & 0 \end{pmatrix}^T e^{\pm ik_{Nx,e}x},$

$$\psi_{Nh+} = \begin{pmatrix} 0, & 0, & 1, & (\dfrac{E_F - E}{\hbar v_x k_{Nx,h} - i\hbar v_y k_{//}}) \end{pmatrix}^T e^{ik_{Nx,h}x},$$

$$\psi_{Ie\pm} = \begin{pmatrix} 1, & (\dfrac{E_F + V_G + E}{\pm\hbar v_x k_{Ix,e} - i\hbar v_y k_{//}}), & 0, & 0 \end{pmatrix}^T e^{\pm ik_{Ix,e}x},$$

$$\psi_{Ih\pm} = \begin{pmatrix} 0, & 0, & 1, & (\dfrac{E_F + V_G - E}{\pm\hbar v_x k_{Ix,h} - i\hbar v_y k_{//}}) \end{pmatrix}^T e^{\pm ik_{Ix,h}x},$$

$$\psi_{Se+} = \begin{pmatrix} 1, & (\dfrac{E_F + U + \Omega}{\hbar v_x k_{Sx,e} - i\hbar v_y k_{//}}), & e^{-i\beta - i\phi}, & e^{-i\beta - i\phi}(\dfrac{E_F + U + \Omega}{\hbar v_x k_{Sx,e} - i\hbar v_y k_{//}}) \end{pmatrix}^T e^{ik_{Sx,e}x},$$

$$\psi_{Sh-} = \begin{pmatrix} 1, & (\dfrac{E_F + U - \Omega}{-\hbar v_x k_{Nx,h} - i\hbar v_y k_{//}}), & e^{i\beta - i\phi}, & e^{i\beta - i\phi}(\dfrac{E_F + U - \Omega}{-\hbar v_x k_{Nx,h} - i\hbar v_y k_{//}}) \end{pmatrix}^T e^{-ik_{Sx,h}x}$$

with $k_{Nx,e} = \dfrac{(E_F + E)\cos[\theta]}{\hbar\sqrt{v_x^2 \cos^2[\theta] + v_y^2 \sin^2[\theta]}}$ , $k_{Nx,h} = \dfrac{(E_F - E)\cos[\theta_A]}{\hbar\sqrt{v_x^2 \cos^2[\theta_A] + v_y^2 \sin^2[\theta_A]}}$ ,

$k_{Ix,e} = \dfrac{(E_F + V_G + E)\cos[\theta_I]}{\hbar\sqrt{v_x^2 \cos^2[\theta_I] + v_y^2 \sin^2[\theta_I]}}$ , $k_{Ix,h} = \dfrac{(E_F + V_G - E)\cos[\theta_{IA}]}{\hbar\sqrt{v_x^2 \cos^2[\theta_{IA}] + v_y^2 \sin^2[\theta_{IA}]}}$ ,

$k_{Sx,e} = \dfrac{(E_F + U + \Omega)\cos[\theta_{SI}]}{\hbar\sqrt{v_x^2 \cos^2[\theta_S] + v_y^2 \sin^2[\theta_S]}}$ , $k_{Sx,h} = \dfrac{(E_F + U - \Omega)\cos[\theta_{SA}]}{\hbar\sqrt{v_x^2 \cos^2[\theta_{SA}] + v_y^2 \sin^2[\theta_{SA}]}}$

and $\Omega = \sqrt{E^2 - \Delta^2}$ and $e^{\pm i\beta} = (E \pm \sqrt{E^2 - \Delta^2})/|\Delta|$ .

$$(9)$$

We can easily calculate the angles of incidences as a function of the injected angle, $\theta$, for quasielectrons and quasiholes in the NG-, I- and SG- regions through the formalism which is related to the conservation of the parallel component $k_{//}$, as given by

$k_{//} = k_{Nx,e} \sin[\theta]/\cos[\theta] = k_{Nx,h} \sin[\theta_A]/\cos[\theta_A] = k_{Ix,e} \sin[\theta_I]/\cos[\theta_I] =$

$k_{Ix,h} \sin[\theta_{IA}]/\cos[\theta_{IA}] = k_{Sx,e} \sin[\theta_S]/\cos[\theta_S] = k_{Sx,h} \sin[\theta_{SA}]/\cos[\theta_{SA}]$

$$(10)$$

The coefficients a, b, l, m, p, q, c, and d can be calculated by using the boundary conditions at x=0 and x=d, as given by

$\psi(x < 0, y)_{x=0} = \psi(0 < x < d, y)_{x=0}$, and $\psi(0 < x < d, y)_{x=d} = \psi(d < x, y)_{x=d}$ .

$$(11)$$

After substituting the wave function in eqn.(9) into the boundary condition in eqn.(11), we can thus determine the Andreev reflection amplitude, a, and the normal reflection amplitude, b. By setting $V_G \to \infty$ and $d \to 0$ for NG/I/SG junction for the case of the thin barrier limit, we have defined $Z \sim dV_G/\hbar v_F$ denoted as the barrier strength. The Andreev and the normal reflection amplitudes are given by

$$a = \frac{-4(A_{e1} - A_{e2})(C_1 - C_2)e^{-i\phi}e^{2iZ_x}e^{i\beta}}{m_1 + m_2 + 2e^{2iZ_x}(m_3 + m_4)} ,$$

and

$$b = \frac{b_1 + b_2 + 2e^{2iZ_x}(b_3 + b_4)}{m_1 + m_2 + 2e^{2iZ_x}(m_3 + m_4)} ,$$

respectively, where

$m_1 = e^{4iZ_x}(-1 + e^{2i\beta})(1 + A_{e2})(1 + A_h)(-1 + C_1)(-1 + C_2)$ ,

$m_2 = (-1 + e^{2i\beta})(-1 + A_{e2})(-1 + A_h)(1 + C_1)(1 + C_2)$ ,

$m_3 = -\left\{(1 + e^{2i\beta})(C_1 - C_2)A_h\right\} + \left\{(-1 + e^{2i\beta})(-1 + C_1C_2)\right\}$ ,

$m_4 = A_{e2}[\left\{(1 + e^{2i\beta})(C_1 - C_2)\right\} - \left\{(-1 + e^{2i\beta})(-1 + C_1C_2)A_h\right\}]$ ,

and

$$b_1 = -e^{4iZ_x}(-1+e^{2i\beta})(1+A_{el})(1+A_h)(-1+C_1)(-1+C_2),$$

$$b_2 = -(-1+e^{2i\beta})(-1+A_{el})(-1+A_h)(1+C_1)(1+C_2),$$

$$b_3 = (1+e^{2i\beta})(C_1-C_2)A_h - (-1+e^{2i\beta})(-1+C_1C_2),$$

$$b_4 = A_{el}\left\{(1+e^{2i\beta})(-C_1+C_2)+(-1+e^{2i\beta})(-1+C_1C_2)A_h\right\},$$

with $A_{el(2)} = \dfrac{E_F+E}{+(-)\hbar v_x k_{Nx,e} - i\hbar v_y k_{//}}$, $A_h = \dfrac{E_F-E}{\hbar v_x k_{Nx,h} - i\hbar v_y k_{//}}$,

$$C_1 = \frac{E_F+U+\Omega}{\hbar v_x k_{Sx,e} - i\hbar v_y k_{//}}, \qquad C_2 = \frac{E_F+U-\Omega}{-\hbar v_x k_{Sx,h} - i\hbar v_y k_{//}}$$

and $Z_x = Z(\dfrac{v_F}{v_x})$.

$$(12)$$

## 2.3 Formulism of the tunneling conductance and the Andreev reflection probability amplitude

We can then calculate the conductance of the junction using the Blonder–Thinkham–Klapwijk (BTK) formalism [10]. The dimensionless conductance in the x-direction is, therefore, given by

$$G_x \sim \int_0^{\theta_c} d\theta \cos\theta (1+(\frac{\sqrt{v_x^2\cos^2[\theta]+v_y^2\sin^2[\theta]}}{\sqrt{v_x^2\cos^2[\theta_A]+v_y^2\sin^2[\theta_A]}})\frac{\cos\theta_A}{\cos\theta}|a(\theta)|^2 - |b(\theta)|^2),$$

where $\theta_C = \cot^{-1}\left(\frac{v_x}{v_y}\sqrt{\left(\frac{E_F+E}{E_F-E}\right)^2-1}\right).$ (13)

The angle-dependent Andreev reflection probability amplitude is also defined as

$$A_x(\theta) \sim (\frac{\sqrt{v_x^2\cos^2[\theta]+v_y^2\sin^2[\theta]}}{\sqrt{v_x^2\cos^2[\theta_A]+v_y^2\sin^2[\theta_A]}})\frac{\cos\theta_A}{\cos\theta}|a(\theta)|^2 \quad (14)$$

In the case of $G_y$ and $A_y(\theta)$, the conductance and the Andreev reflection probability amplitude are related to the current $I_y$. They can easily be determined by interchange $v_x \leftrightarrow v_y$ in the previous formulae, ie., $G_y \sim G_x(v_x \leftrightarrow v_y)$ and $A_y \sim A_x(v_x \leftrightarrow v_y)$.

## 3.    Result and discussion

We first consider the angle-dependent Andreev probability amplitudes $A_x(\theta)$ and $A_y(\theta)$ using eqn.(14) for the various values of strain S=0, 0.2, 0.22 and 0.2288. As we mentioned in the previous section (see Fig. 1b), the transition point between gapless graphene to gapped graphene is at strain of $S_C$=0.228855. In this section we need value of strain near the critical value $S_C$ to show the effect of the highly asymmetric velocity $v_y \gg v_x$ on the Andreev reflection. The Andreev probability amplitudes are studied for the case of $U = 5\Delta, E_F = 0.5\Delta, eV = 0,$ and $Z = 0$ as seen in Figs.3a-3b. Our focus is to show the effect of the asymmetric-velocity fermions, which form the Cooper pairs in the system, on the Andreev reflection at the NG/SG interface. We find that in case of the current in the x-direction (see Fig.3a), $A_x(\theta)$ is suppressed by strain for large angle of incidence. For all values of strain, it is smaller than that in the unstrained graphene system (Strain=0). When increasing strain approaching $S_C \sim 0.228855$ ($v_y/v_x \sim$ very large), $A_x(\theta)$ is suppressed, except for the normal incidence. This is to say that it allows only the current at $\theta = 0$ which yields $A_x = 1$, showing the presence of the Klein tunneling [5] due to relativistic fermions with zero mass. Increasing strain approaching $S_C$, $A_y(\theta)$ is almost~1 for all angles of incidence, which is rather different from the current in the x-direction (see Fig.3b). This novel behaviour, the direction-dependent Andreev reflection $A_x(\theta) \neq A_y(\theta)$, results from the asymmetric massless fermions with $v_x \neq v_y$ in strained graphene system, in contrast to the Andreev reflection of the symmetric massless fermions $v_x = v_y$ in the unstrained graphene NG/SG system with yielding $A_x(\theta) = A_y(\theta)$ [8].

Based on eqn.(13),the tunneling conductances $G_x$ and $G_y$ as a function of the biased voltage V are first studied in case of Z=0, NG/SG junctions. The parameters, $U = 5\Delta$ set as weakly doped graphene in the SG region and $E_F = 0.5\Delta$, are assumed to show the effect of the specular Andreev reflection when $E_F < eV$ on the conductances. We first consider $G_x$ for strain of S=0, 0.2, and 0.2288 (see Fig.4a). The curve, for strain=0, is to show the conductance due to the direction-independent velocity fermions in the unstrained graphene NG/SG conventional junction, as predicted previously in refs. 11-13. In this direction, increasing strain approaching $S_C$ leads to the conductance vanishing. Strain$\rightarrow$Sc gives rise to $v_x$~very small. Previously obtained in eqn.(5), we have $v_x$(strain=0, 0.2 and 0.2288)= $v_F$, $0.342292v_F$ and $0.014824v_F$, respectively. As very different from $G_x$, the conductance $G_y$ seen in Fig.4b increases with increasing strain for all eV. Remarkably, for strain=0.2288 a similar perfect current switch at eV=$E_F$, the transition point between specular Andreev reflection and the retro Andreev reflection is observed in this junction. This is very different from that in the unstrained case (strain=0, see refs.11-13) and it may be applicable for nanoswicth devices. The strain dependence of velocity $v_y$(strain=0, 0.2 ,and 0.2288)= $v_F$ , $1.08075v_F$, and $1.0928 \; v_F$, respectively. The velocity ratio $v_y/v_x$=73% for strain =0.2288. As a conclusion, rising velocity ratio $v_y/v_x \gg 1$ by increasing strain approaching $S_C$ results in better current switch for the conductance in the y-direction at eV=$E_F$.

We next consider the case of the heavily doped graphene in the SG region $U = 1000\Delta$ for no barrier Z=0 and the case of highly-asymmetric-velocity particles ($v_y/v_x$=73% for strain =0.2288), as seen in Figs.5a-5b. The conductances in NG/SG junction are calculated as a function of the biased voltage V for various values of $E_F$. As a result, the behavior of the conductance $G_y$ is rather different from that of the conductance $G_x$. In case of the conductance $G_y$, for small $E_F$=$0.1\Delta$, $0.5\Delta$, $\Delta$ and $1.5\Delta$, the behavior of conductance is similar to that of $G_y$ for the case of weakly doped graphene (U~small). For the large Fermi energy $E_F$~$1000\Delta$, the conductance is similar to the case of the unstrained graphene-based NG/SG junction [8, 11-13], as is strain-independent. But in the case of $G_x$, the conductance is rather small. The conductanc peak due to the Andreev resonance is found for the large $E_F$~$1000\Delta$.

In Figs. 6-7, the conductances are plotted as a function of the barrier strength $Z \sim V_G d / \hbar v_F$ in NG/I/SG junctions, for various values of strain 0, 0.20, 0.22 and 0.2288. In Fig6a, the conductance $G_y$ is first investigated. We set U=0, $E_F$=100$\Delta$, and eV=0, as the case of zero biased voltage and as the case of the non-Fermi-energy mismatch in NG and SG. For strain=0, we have the same curve as that in refs.12-13 of the unstrained case. Interestingly, when increasing strain approaching $S_C$, the current flows through the junction with G~2 as if there was no barrier. This is to show that when $v_y/v_x \gg 1$, the general effect of the gate voltage is destroyed by the highly-asymmetric-velocity effect for $G_y$. In Fig6b, we take into account the effect of the Fermi-energy mismatch U=900$\Delta$. Increasing U decreases the amplitude of $G_y$ [11-13]. We find that increasing strain approaching $S_C$ also destroys the effect of gate voltage, like the behavior of the case for U~0. Let us next consider the conductance $G_x$ as a function of the barrier strength Z, which is numerically shown in Fig7. For U=0 (case of non-Fermi-energy mismatch), $E_F$=100$\Delta$, and eV=0, the anomalous conductance oscillation with very high frequency is found when strain is of 0.2288 ($v_y/v_x$=73%). The increasing frequency of the oscillation in $G_x$ can be described via eqn.(12). The anomalous oscillation is due to the term of "exp[iZ($v_F/v_x$)]". This is to show straightforwardly that the frequency related to the term of "exp[iZ($v_F/v_x$)]" is proportional to ~1/$v_x$. The small value of $v_x$=0.014824$v_F$ for strain of 0.2288 gives rise to the high frequency. However, this anomalous behavior, which is rather different from the case of the unstrained NG/I/SG junction [11-13], is not observed when U is very large (see Fig.7b for the case of U=900$\Delta$).

## 4.      Summary and conclusion

We have investigated the conductances in strained graphene-based NG/I/SG junctions where graphene sheet is strained in the zigzag direction. This work studied the conductance based on the BTK formalism and based on the assumption that by depositing conventional superconductor on the top of the strained graphene sheet, graphene can be a superconductor with the Cooper pairs formed by the asymmetric Weyl-Dirac electrons, instead of the symmetric Weyl-Dirac electrons in the case of unstrained graphene system. Strain in the zigzag direction gives rise to the highly-asymmetric-velocity massless fermions, asymmetric Weyl–Dirac fermions with $v_x \ll v_y$, as the carriers of the system when strain approaches the critical point, the point of the transition between gapless and gapped graphene [21-22]. In our model, we used the geometrically deformed graphene based on the model of ref.21, leading to the critical strain $S_C$~0.228855. In this work, we focused on the effect of strain near the $S_C$ which causes the strong effect of the highly-asymmetric-velocity fermions on the Andreev reflection and the conductances of the junctions. The currents were investigated for the two cases which are parallel and perpendicular to the direction of strain. As a result, because we have taken into account the effect of asymmetric velocity $v_x \neq v_y$ resulting from strain on the superconducting transport property, we found a novel feature of the Andreev reflection and the tunneling conductance which have not been predicted in the previously unstrained graphene-based NG/I/SG junctions [8, 11-13]. All of our theoretically predicted results should be experimentally testable.

**Figure captions**

**Figure 1** shows (a) the geometry of graphene structure where the hoping energies $t_1 = t_2 \neq t_3$ related to the displacement vectors of the nearest neighbor atoms $\vec{\sigma}_1, \vec{\sigma}_2$ and $\vec{\sigma}_3$, respectively and (b) the velocity ratio $v_y/v_x$ for graphene sheet being strained in the zigzag or x direction and the armchair or y direction. The highly asymmetric velocity is found only the case where graphene is strained in the zigzag direction, $v_y/v_x \to \infty$ for strain $\to S_C$.

**Figure 2** shows the present models of strained graphene-based NG/I/SG junctions where graphene is strained in the **zigzag direction** for (a) the case of current $I_x$ parallel to the direction of strain and (b) the case of current $I_y$ perpendicular to the direction of strain**.** The two junctions are biased by the voltage V and the gate potential applied in the barrier is denoted as $V_G$. The injected angle of quasiparticles at the interface of the NG/I/SG junction is denoted as $\theta$.

**Figure 3** shows the effect of strain on angle-dependent Andreev reflection probability amplitude in NG/I/SG junctions, where we set Z=0, $U = 5\Delta$, $E_F = 0.5\Delta$ and eV=0, (a) for $A_x$ due to the current in the x-direction and (b) for $A_y$ due to the current in the y-direction.

**Figure 4** shows the effect of strain on the conductances as a function of the biased voltage V in NG/I/SG junction for Z=0, $E_F = 0.5\Delta$ and $U = 5\Delta$, (a) for conductance $G_x$ related to the current in the x-direction and (b) for conductance $G_y$ related to current in the y-direction. Strain increases current in the y-direction but decreases current in the x-direction.

**Figure 5** shows the conductances as a function of the biased voltage V in NG/I/SG junction for Z=0, strain of 0.2288, and $U = 1000\Delta$, (a) for conductance $G_y$ with various values of $E_F$ and (b) for conductance $G_x$ with various values of $E_F$.

**Figure 6** shows the conductance $G_y$ as a function of the barrier strength Z in NG/I/SG junction for eV=0, and $E_F = 100\Delta$, (a) for $U = 0$ (case of $E_{FS}/E_{FN}=1$) with various values of strain and (b) for $U = 900\Delta$ (case of $E_{FS}/E_{FN}=10$) with various values of strain.

**Figure 7** shows the conductance $G_x$ as a function of the barrier strength Z in NG/I/SG junction for eV=0, and $E_F = 100\Delta$, (a) for $U = 0$ (case of $E_{FS}/E_{FN}=1$) with various values of strain and (b) for $U = 900\Delta$ (case of $E_{FS}/E_{FN}=10$) with various values of strain.

**(1a)**

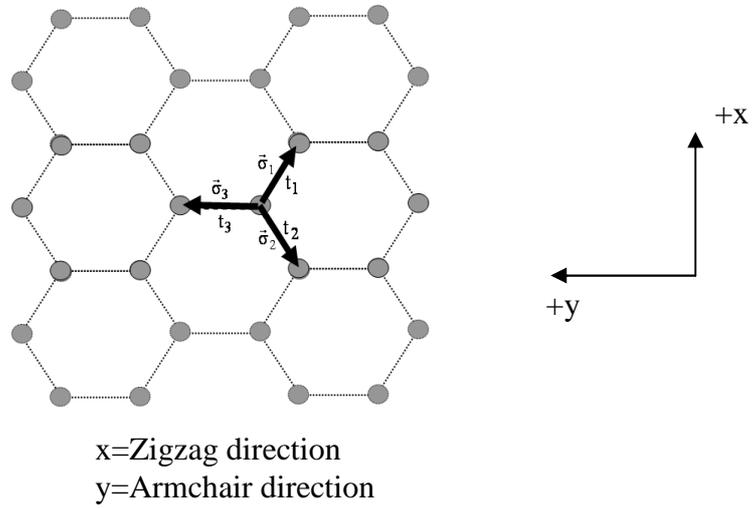

x=Zigzag direction
y=Armchair direction

**(1b)**

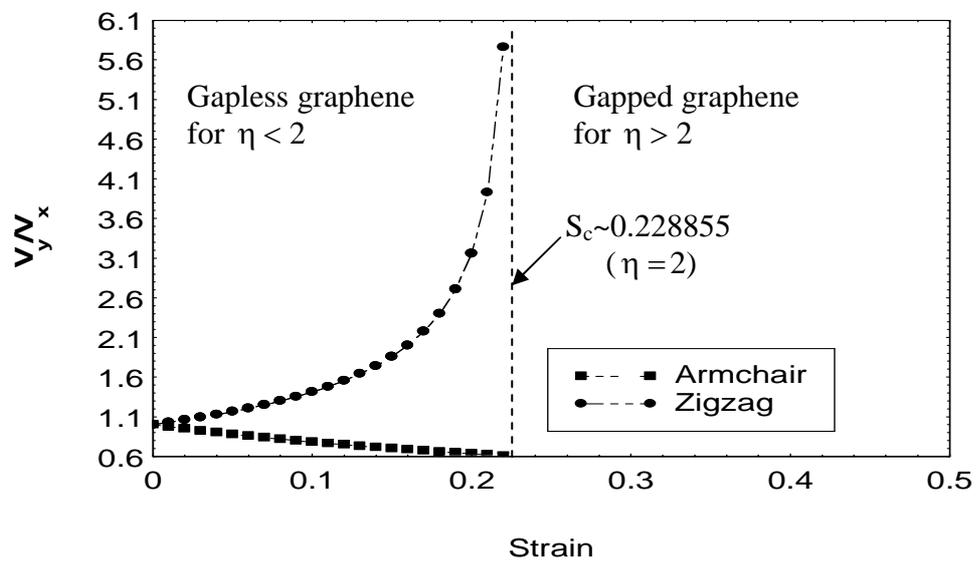

**Figure 1**

(2a)

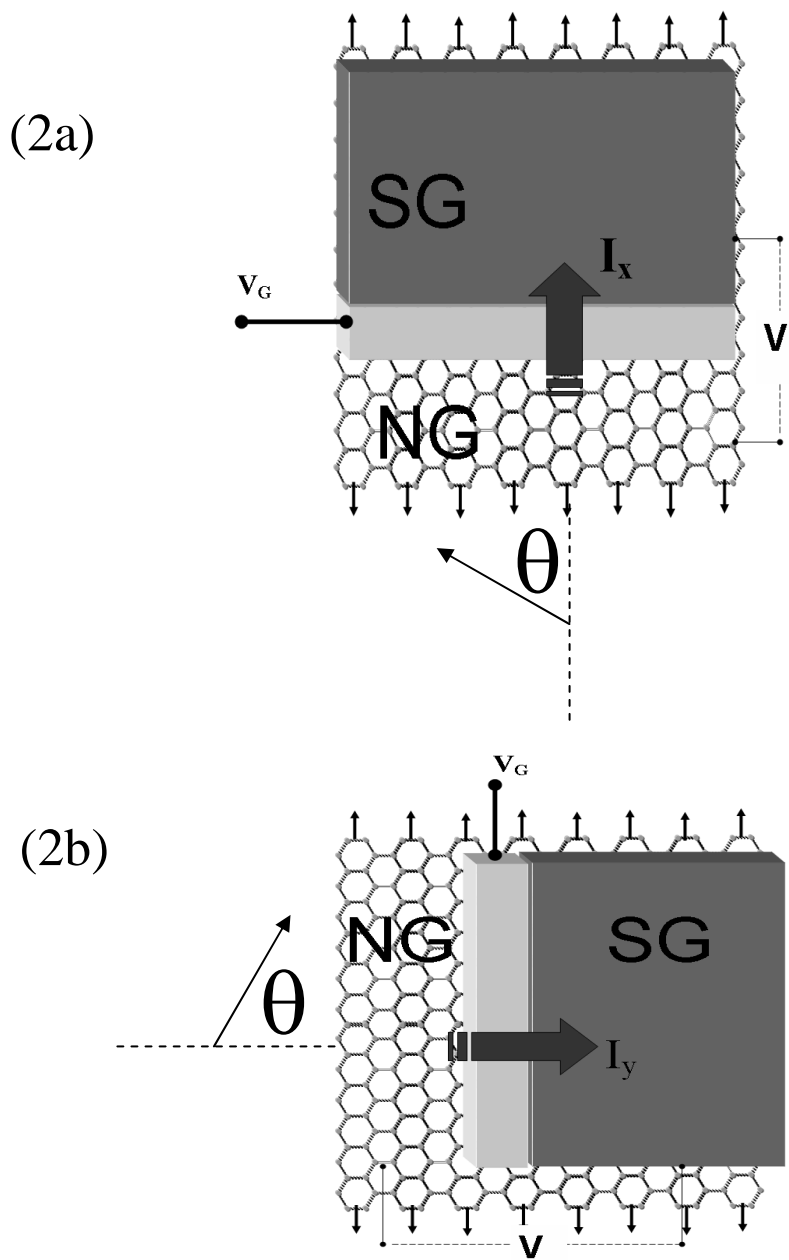

**Figure 2**

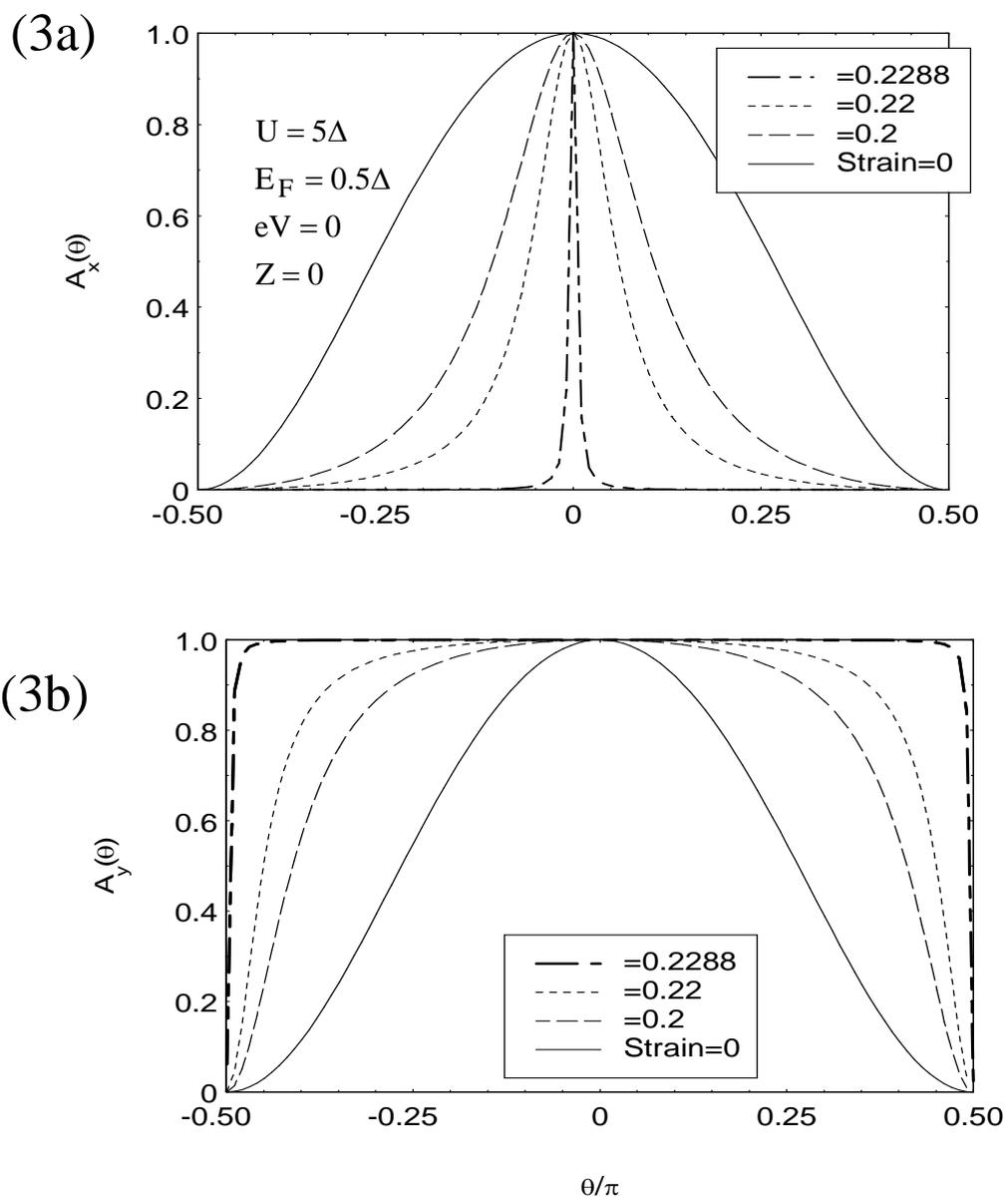

**Figure 3**

(4a)

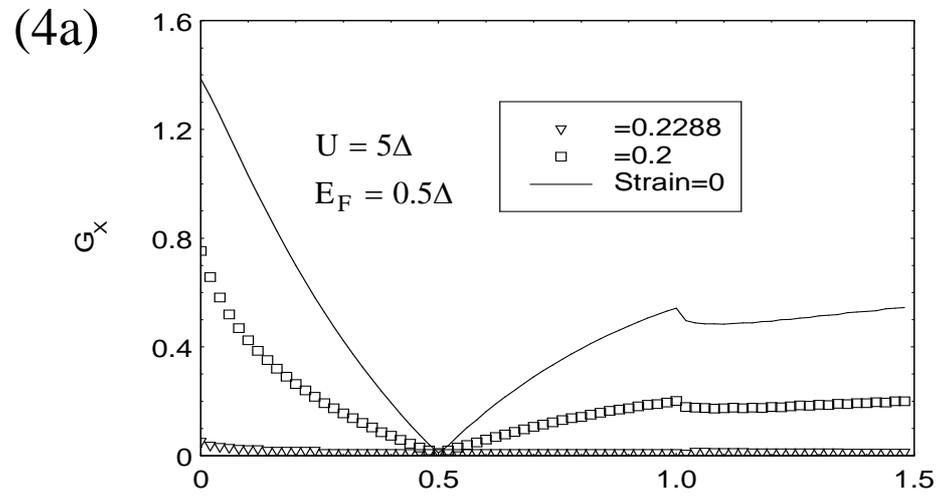

(4b)

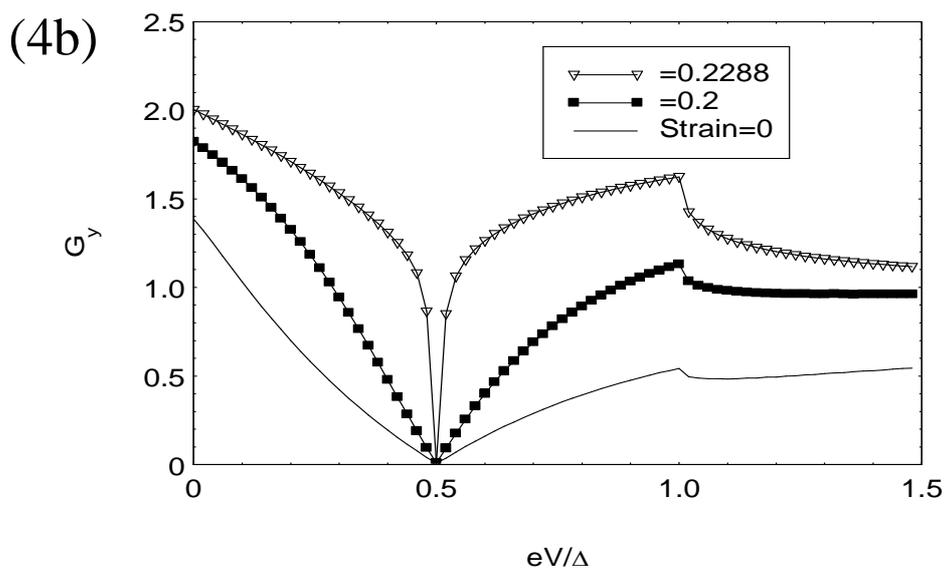

**Figure 4**

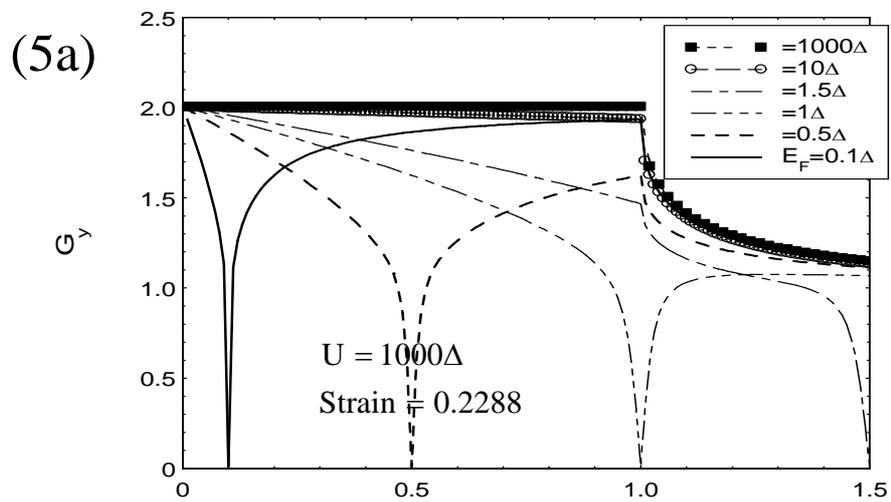

(5a)

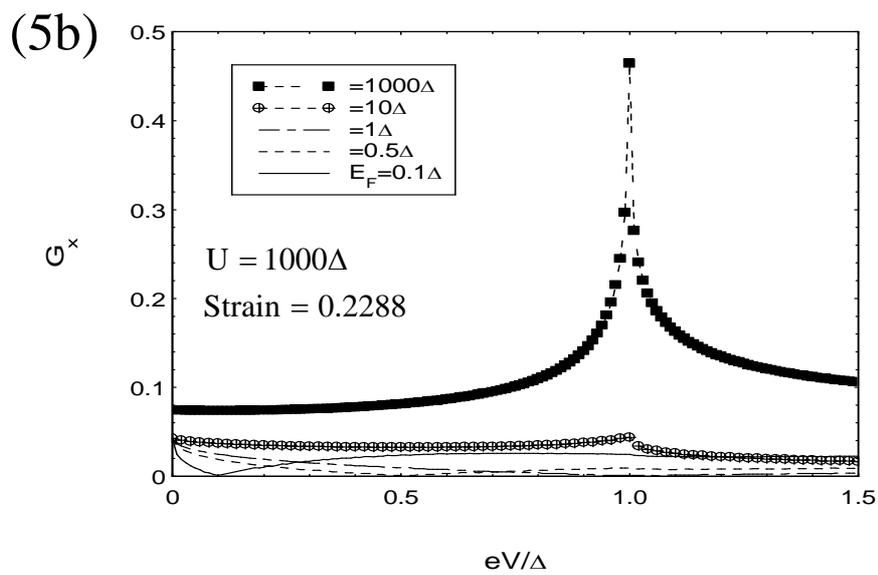

(5b)

**Figure 5**

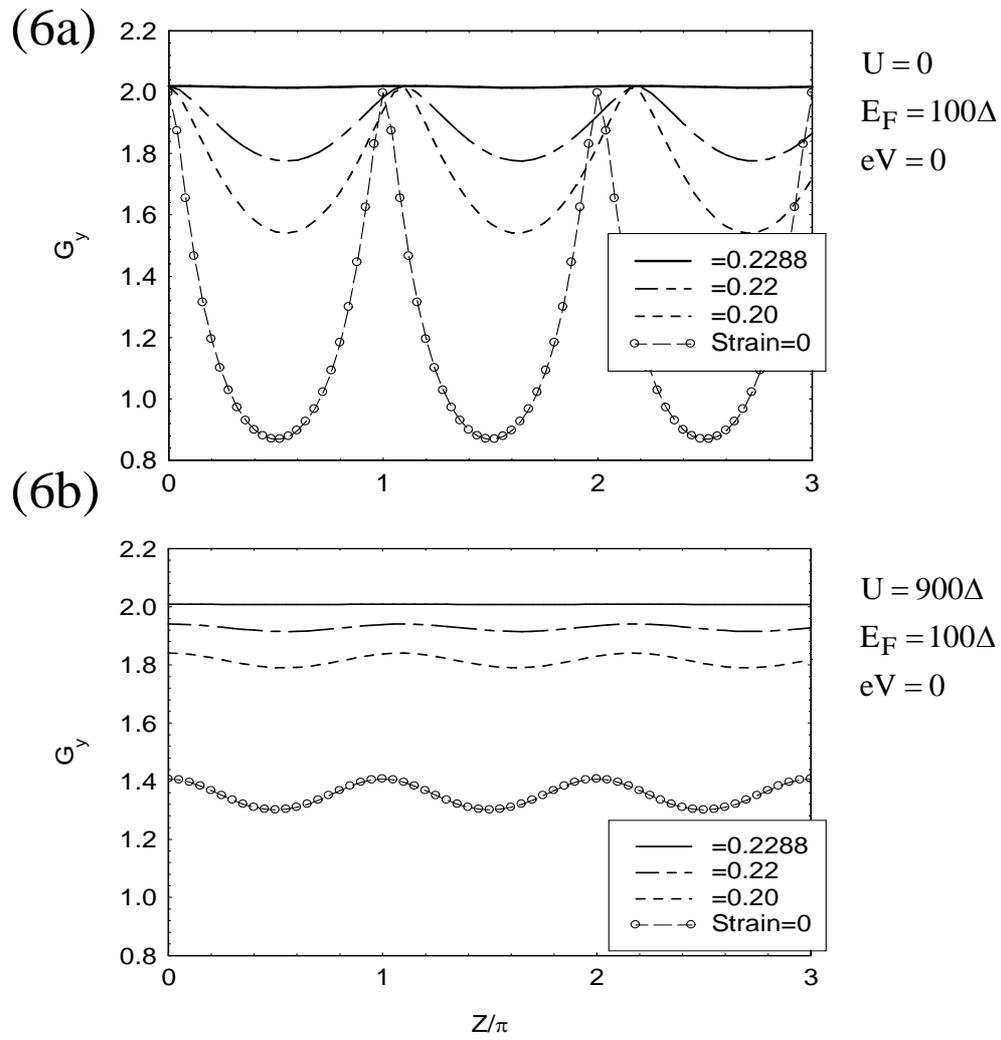

(6a)

$G_y$

U = 0
$E_F = 100\Delta$
eV = 0

Legend:
——— =0.2288
– – – =0.22
- - - =0.20
-○-○- Strain=0

(6b)

$G_y$

Z/π

U = 900$\Delta$
$E_F = 100\Delta$
eV = 0

Legend:
——— =0.2288
– – – =0.22
- - - =0.20
-○-○- Strain=0

**Figure 6**

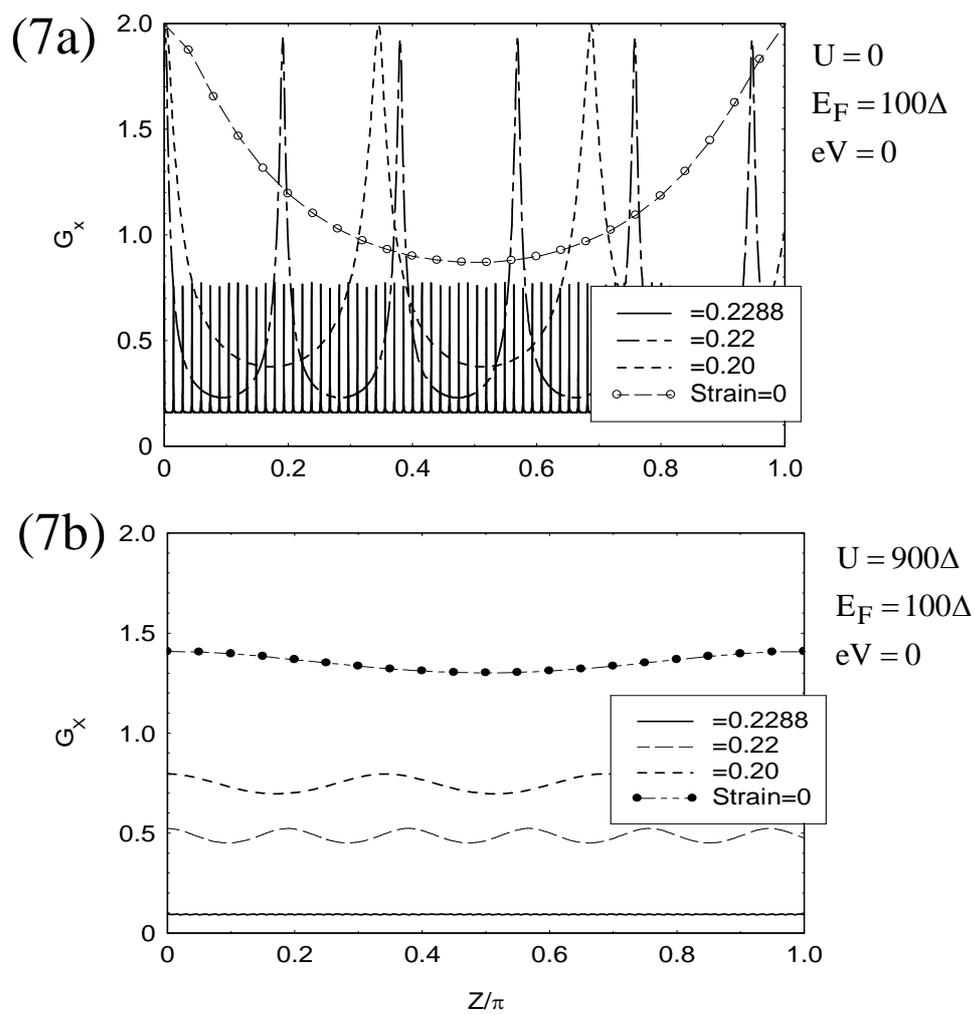

(7a)

U = 0
$E_F = 100\Delta$
eV = 0

$G_x$

2.0
1.5
1.0
0.5

| | |
|---|---|
| —— | =0.2288 |
| – – | =0.22 |
| - - - | =0.20 |
| –○– | Strain=0 |

0    0.2    0.4    0.6    0.8    1.0

(7b)

U = 900Δ
$E_F = 100\Delta$
eV = 0

$G_x$

2.0
1.5
1.0
0.5

| | |
|---|---|
| —— | =0.2288 |
| – – | =0.22 |
| - - - | =0.20 |
| –●– | Strain=0 |

0    0.2    0.4    0.6    0.8    1.0

Z/π

**Figure 7**